\documentstyle[12pt,psfig]{article}
\def\ii{\'{\char'20}}
\begin{document}
\newcommand{\app}[1]{\section{#1}\renewcommand{\theequation}
        {\mbox{\Alph{section}.\arabic{equation}}}\setcounter{equation}{0}}
\newcommand{\pol}{\left(\frac 1 {s+1/2}+\ln R^2\right)}
\newcommand{\inu}{\int\limits_0^{\infty}d\nu\,\,}
\newcommand{\nen}{1+e^{2\pi\nu}}
\newcommand{\nr}{\nu^2-m^2R^2}
\newcommand{\mrnu}{\left(\frac{mR}{\nu}\right)^2}
\newcommand{\ep}{\epsilon}
\newcommand{\fr}{\frac}
\newcommand{\reals}{\mbox{${\rm I\!R }$}}
\newcommand{\nats}{\mbox{${\rm I\!N }$}}
\newcommand{\intgs}{\mbox{${\rm Z\!\!Z }$}}
\newcommand{\cam}{{\cal M}}
\newcommand{\caz}{{\cal Z}}
\newcommand{\cao}{{\cal O}}
\newcommand{\cac}{{\cal C}}
\newcommand{\aaa}{\int\limits_{mR}^{\infty}dk\,\,}
\newcommand{\bbb}{\left[\left(\frac k R\right)^2-m^2\right]^{-s}}
\newcommand{\ccc}{\frac{\partial}{\partial k}}
\newcommand{\fff}{\frac{\partial}{\partial z}}
\newcommand{\iikma}{\aaa \bbb \ccc}
\newcommand{\ddd}{\int\limits_{mR/\nu}^{\infty}dz\,\,}
\newcommand{\eee}{\left[\left(\frac{z\nu} R\right)^2-m^2\right]^{-s}}
\newcommand{\lll}{\frac{(-1)^j}{j!}}
\newcommand{\iinma}{\ddd\eee\fff}
\newcommand{\cah}{{\cal H}}
\newcommand{\nn}{\nonumber}
\renewcommand{\theequation}{\mbox{\arabic{section}.\arabic{equation}}}
\newcommand{\komplex}{\mbox{${\rm I\!\!\!C }$}}
\newcommand{\sip}{\frac{\sin (\pi s)}{\pi}}
\newcommand{\numr}{\left(\frac{\nu}{mR}\right)^2}
\newcommand{\mzs}{m^{-2s}}
\newcommand{\rzs}{R^{2s}}
\newcommand{\abl}{\partial}
\newcommand{\g}{\Gamma\left(}
\newcommand{\ikma}{\int\limits_{\gamma}\frac{dk}
                     {2\pi i}\,\,(k^2+m^2)^{-s}\frac{\partial}{\partial k}}
\newcommand{\ead}{e_{\alpha}(D)}
\newcommand{\sual}{\sum_{\alpha =1}^{D-2}}
\newcommand{\sulnu}{\sum_{l=0}^{\infty}}
\newcommand{\sujnu}{\sum_{j=0}^{\infty}}
\newcommand{\suani}{\sum_{a=0}^i}
\newcommand{\suanzi}{\sum_{a=0}^{2i}}
\newcommand{\zend}{\zeta_D^{\nu}}
\newcommand{\amed}{A_{-1}^{\nu ,D}(s)}
\renewcommand{\and}{A_{0}^{\nu ,D}(s)}
\newcommand{\aid}{A_{i}^{\nu ,D}(s)}
\def\beq{\begin{eqnarray}}
\def\eeq{\end{eqnarray}}
\begin{titlepage}
\title{\begin{flushright}
{\normalsize UB-ECM-PF 96/14 }
\end{flushright}
\vspace{3mm}
{\Large \bf Casimir energies for massive fields in the bag}}
\author{M. Bordag,$^{a,}$\thanks{E-mail address:
bordag@qft.physik.uni-leipzig.d400.de} \ 
E. Elizalde,$^{b,c,}$\thanks{E-mail address: eli@zeta.ecm.ub.es} \
K. Kirsten$^{a,}$\thanks{E-mail address: kirsten@tph100.physik.uni-leipzig.de} \
and S. Leseduarte$^{c,}$\thanks{E-mail address: lese@zeta.ecm.ub.es}\\ 
\mbox{}\\
$^{a}$ Universit{\"a}t Leipzig, Institut f{\"u}r Theoretische Physik,\\
Augustusplatz 10, 04109 Leipzig, Germany\\
$^{b}$ Centre d'Estudis Avan\c cats de Blanes, CSIC, \\ Cam\'{\i} de Santa
B\`arbara, 17300 Blanes, Spain\\ 
 $^{c}$ Departament d'ECM and IFAE, Facultat de F{\ii}sica,\\
Universitat de Barcelona, Av. Diagonal 647, \\   08028 Barcelona,
Spain}
\vspace*{-1mm}
\maketitle
\vspace*{-2mm}

\begin{abstract}
The Casimir energy corresponding to a massive scalar field with 
Dirichlet boundary conditions on a spherical bag
is obtained. The field is considered, separately,
inside and outside the bag. The renormalization procedure that is
necessary to apply in each situation is studied in detail, in particular
the differences occurring with respect to the case when the field occupies
the whole space. The final result contains several constants that
experience renormalization and can be determined only experimentally.
The non-trivial finite parts that appear in the massive case are found
exactly, providing a precise determination of the complete, renormalized 
zero-point energy for the first time.
\end{abstract}

\vfill \noindent PACS: 02.30.+g, 02.40.+m \\
hep-th/9608071\\

\end{titlepage}
\section{Introduction}
\setcounter{equation}{0}
Calculations of the Casimir energy in spherically symmetric situations
have attracted the interest of physicists for well over forty years 
now. This is not strange, since in
many contexts the inclusion of quantum fluctuations about semi-classical
configurations turns out to be essential. 
Historically a first far reaching
idea involving vacuum energies originated with Casimir himself.
He proposed that the force
stabilizing a classical electron model arises from the zero-point
energy of the electromagnetic field within and without a perfectly
conducting spherical shell \cite{casimir56}. Having found an
attractive force between parallel plates due to the vacuum energy
\cite{casimir48}, the hope was that the same would occur for the
spherically symmetric situation\footnote{This could be expected because the Casimir force between plates was shown to be the same as the force resulting from the retarded (always attractive) van der Waals forces between the atoms of the plates.}. Unfortunately, as Boyer 
\cite{boyer68} first showed, for this geometry the stress is
repulsive \cite{baldup78,mil78}. Nowadays it is known that the 
Casimir energy depends strongly on the geometry of the space-time
and on the boundary conditions imposed. This is still a very active field
of research
because a satisfactory understanding of the behaviour has not yet been 
found. 
For a number of results obtained in the last five years
 see for instance
\cite{eorbz}-\cite{greiner}.

More recently the zero-point energy has got considerable attention 
in the context
of the bag model \cite{cho74}-\cite{has78} and chiral bag model
\cite{vep90}-\cite{brow84}.
In those models the quarks and gluons are free inside the bag, but are
absolutely confined to the interior of the bag. 
The sum of the mesonic, valence quark and vacuum quark contributions to
the baryonic number have been found to be independent of the bag radius
and of the pion field strength, being the vacuum quark contributions
---which are analogue to the Casimir effect in QED--- essential in
the calculation of baryonic observables. The issue of regularization
is certainly non-trivial. Under some circumstances,
different regularization procedures can yield different results and
real physical problems arise in the calculation of quark vacuum
 contributions to some barionic observables, as the energy itself.


For massless fermions
the zero-point energy was considered in \cite{milton80,milton83}.
 The massless fermionic field inside and outside the spherical bag
was analysed  in \cite{milton83}. In the last case, a cancellation of
divergences between the inner and outer spaces occurs and finite 
zero-point energies are found. Considering only the inner space, divergences
appear and it is necessary to introduce contact terms and perform a 
renormalization of their coupling. Results for the massive fermionic 
fields contain new ultraviolet-divergent terms in addition to those 
occurring in the massless case, as has been discussed
in \cite{baacke83}. Further considerations, especially on the
 renormalization procedure
necessary in order to carry out
these calculations, and on its precise interpretation, can be found in
\cite{andreas}.

In nearly all of the mentioned works the authors have used a Green's
function
approach in order to calculate the zero-point energy. An exception is Ref.
\cite{andreas}, where, in the general setting of an ultrastatic spacetime
with or without boundaries, a systematic procedure which makes use of
 zeta function
regularization was developed. In this approach, a know\-ledge of the zeta
function of the operator associated with the field equation together
with (eventually) some appropriate
 boundary conditions is needed. Recently, a detailed
description of how to obtain the zeta function for a massive scalar field
inside a ball satisfying Dirichlet or Robin boundary conditions has been
given by some of the
authors \cite{bk,bek} of the present work. An analytical continuation
to the whole complex plane has been obtained and then applied to find an
arbitrary number of heat-kernel coefficients.
In the ensuing Refs.
\cite{begk,bdk} the functional
determinant has been considered too
and, furthermore, the method
has been also applied to spinors \cite{abdk} and p-forms \cite{eli1,eli2}.
For an alternative approach involving scalars and spinors see also the 
developments in \cite{dowker}.
All the above considerations yield purely analytical and quite 
explicit formulas. In order to actually
obtain  values for the Casimir energy, however,
a numerical integration had to be performed.
 This has been  done in different cases, in particular
 for the massless scalar and electromagnetic fields 
\cite{rom}, partly reobtaining previous
results.

Here we will consider the massive scalar field
inside and outside
the bag, separately. We will discuss in detail
the renormalization procedure which is
 necessary to apply in
this situation and, specifically,
the differences occurring with respect to the case
 when one assumes that the field occupies the whole space.
The final result 
for the zero-point energy
contains several constants which experience renormalization
and whose physical values can be determined only experimentally.
However, for the massive field
---as is clear from dimensional grounds---
 non-trivial finite parts which depend on an adimensional variable
involving the mass are present, and we will, in this case, be able to 
find here for the first time the complete, renormalized
zero-point energy.

The organization of the paper is as follows. 
In Sect.~2 we describe in detail the models considered and the 
regularization and renormalization procedure employed. 
In Sect.~3 we summarize
 briefly the formulas that are needed in the subsequent study of
 the zeta function of the problem at hand. We shall start
with the scalar field inside the ball. Some additional considerations are
 necessary because the representation given in the previous articles
\cite{bk,bek} was only applicable when the mass of the field is $m\leq 1$.
Here we will derive formulas which are valid for arbitrary mass and
very useful for numerical
evaluations. All divergences and finite parts of the zero-point energy
are calculated and
its renormalization is performed with some care. The explicit dependence
of the finite
part in the mass is determined numerically. Afterwards, the
exterior space is considered and a corresponding analysis is performed
for this situation.
 Adding
up both contributions, we  see clearly
how the divergences cancel among themselves as well as the
influence of this cancellation on the
compulsory renormalization process.
Sect. 4 is devoted to conclusions. 
The appendices contain some hints and technical details that are
 used in the  derivation of the zeta function for the non-zero mass case.

\section{Description of the model and its renormalization}
\setcounter{equation}{0}
The physical system  that we will consider consists of two parts:
\begin{enumerate}
\item A classical system consisting of a spherical
 surface (`bag') of radius $R$. Its energy reads:
\beq
E_{class} = p V + \sigma  S + F R +k +\frac{h}{R} ,\label{n1}
\eeq
where $V=\frac{4}{3}\pi R^3$ and $S=4\pi R^2$ are the volume and 
surface, respectively. This energy is determined by the
 parameters $p =$  pressure, $\sigma =$  surface tension, 
and $F$, $k$, and $h$ which do not have special names. 
\item A quantum field $\varphi (x)$ obeying the equation
\beq (\Box +m^2)\varphi (x) =0, \eeq
as well as boundary conditions on the surface. We choose Dirichlet 
boundary conditions for simplicity. 
The quantum field has a ground state energy
\beq
E_0 = \frac 1 2 \sum_{(k)} \lambda _ {(k)} ^{1/2 }  ,\label{n30}
\eeq
where the $\lambda _ {(k)}$'s are the one-particle energies with the quantum
number $k$. 
\end{enumerate}
For this system we shall consider three models,
 which will behave in a different way.
 These models consist of the classical part given
 by the surface and
\begin{itemize}
\item[ (i)] the quantum field in the interior of the surface,
\item[ (ii)] the quantum field in the exterior of the surface,
\item[ (iii)] the quantum field in both regions together, 
\end{itemize}
respectively.

The ground state energy is divergent and we shall regularize it by
\beq
E_0 = \frac 1 2 \sum_{(k)} \lambda _ {(k)} ^{1/2 -s} \mu^{2s}
 .\qquad \Re s>2. \label{n3}
\eeq
The one particle energies are determined by the eigenvalue equation
\beq
(-\Delta +m^2) \varphi_{(k)} (x) =\lambda_{(k)} \varphi _{(k)} (x)\label{1}
\eeq
with Dirichlet boundary conditions on the surface
\beq \varphi _{(k)} (x)_{\Big |_{ |x|=R}}=0.\eeq
For the field in the interior the meaning of $\lambda_{(k)}$ is obvious:
$(k)=(l,m,n)$,  $\lambda_{(l,m,n)}=\sqrt{j_{l+1/2,n}^2/R^2+m^2}$, \ 
$J_{l+\frac{1}{2}}\left(j_{l+1/2,n}\right)=0$.

For the calculations we use the corresponding zeta function:
\beq\zeta (s)=\sum_{(k)} \lambda_{(k)}^{-s}.\eeq
In the interior region, we have
\beq \zeta_{(int)} (s)=\sum_{l=0}^\infty \sum_{n=0}^\infty (2l+1) 
\lambda_{(l,m,n)}^{-s}.
\label{intzeta}
\eeq
For the exterior zeta function $\zeta_{(ext)} (s)$ we must
take into account that the radial quantum number is continuous. 
We have to subtract the Minkowski space contribution. This procedure 
is well known (see, for example, \cite{borkir}) and  need not be repeated 
 here. 

In the case of the third model we have simply
 to add the interior and the exterior 
zeta functions, namely
\beq
\zeta_{(total)} (s)= \zeta_{(int)} (s)+ \zeta_{(ext)} (s).\label{Z3}
\eeq
Thus, in any case the regularized ground state energy is given by
\beq
E_0^{(mod)} = \frac{1}{2} \zeta _{(mod)} (s-\frac{1}{2}) ~\mu^{2s},\eeq
where $mod$ means the model: $int$, $ext$ or $total$.

The divergent contributions of the ground state energy can
 be found most easily using standard heat kernel expansion
\beq
K (t) = \sum_{(k)} e^{-\lambda_{(k)} t} \sim  \left(
\frac 1 {4\pi t}\right)^{3/2} e^{-tm^2} \sum_{j=0,1/2,1,...}^{\infty}
B_j t^j, \qquad t \to 0^+,\label{n5}
\eeq
by means of 
\beq
\zeta (s) =\frac 1 {\Gamma (s)} \int\limits_0^{\infty} dt\,\,
t^{s-1} K (t)  ,\label{n6}
\eeq
where we have to take into account that in the presense of a boundary
coefficients with half integer numbers are non-zero. As
  is known and can be easily
 found from the above formulas, the first five coefficients $B_i$
($i=0,\frac{1}{2},1,\frac{3}{2},2$) can contribute 
to divergences of the
energy resp. to the pole of the zeta functions. 
The corresponding contributions
to the energy read:
\beq
E_{0(div)} &=&  -\frac{m^4}{64\pi ^2} B_0
\left( \frac 1 {s+\frac{1}{2}}
-\frac 1 2 +\ln \left[\frac{4\mu^2}{m^2}\right]\right)
-\frac{m^3}{24\pi^{3/2}}B_{1/2}\nn\\
& &+\frac{m^2}{32\pi^2} B_1 
\left( \frac 1 {s+\frac{1}{2}}
- 1  +\ln \left[\frac{4\mu^2}{m^2}\right]\right)
+\frac m {16\pi^{3/2}}B_{3/2}\label{n7}\\
& &-\frac 1 {32\pi^2} B_2 
\left( \frac 1 {s+\frac{1}{2}}
- 2 +\ln \left[\frac{4\mu^2}{m^2}\right]\right)  .\nn
\eeq
The heat kernel coefficients are well known (see for instance \cite{kennedy78}) 
and are, for the
interior region,
\beq
B^{(int)}_0 = \frac 4 3 \pi R^3,\quad B^{(int)}_{1/2} =-2\pi^{3/2} R, \quad
B^{(int)}_1 = \frac 8 3 \pi R,\quad
B^{(int)}_{3/2} =-\frac 1 6 \pi^{3/2}, \quad \nn B^{(int)}_2 = \frac {16}{315} \frac
{\pi} R.\nn
\eeq
In the exterior region, we have:
\beq
B^{(ext)}_i &=& B^{(ext)}_i, \qquad i=\frac{1}{2},\frac{3}{2},... ,\\
B^{(ext)}_i &=& - B^{(ext)}_i ,\qquad i=0,1,2,... \label{cancel}.
\eeq

In order to perform the renormalization we choose as
general scheme the following:  all contributions of the heat kernel coefficients 
which can lead to divergences in some regularization have to be subtracted 
by means of a renormalization of the corresponding parameters in the 
classical part 
of the system. 

A remark is in order. The zeta functional regularization used here leaves
the contributions of coefficients with half integer number finite in the limit 
$s\to 0$. This is a specific feature of this regularization and often much 
appreciated. However, in other regularizations, 
as for example the proper time cutoff \cite{andreas}, 
these contributions are divergent. 
Thus, we have in each of the first two models five divergent
contributions. In the third model we note that,  in accordance with (\ref{Z3}), 
\[E_{0(total)}^{(div)}=E_{(div)}^{(int)}+E_{(div)}^{(ext)},\]
and due to the known cancellation of divergent contributions,
which is in fact due to (\ref{cancel}), only two of them remain.

As  physical system we consider the classical part and the ground state 
energy of the quantum field together and write for the complete energy
\beq
E=E_{(class)}+E_0.
\eeq
In this context the renormalization can be achieved by shifting the parameters 
in $E_{(class)}$ by an amount which cancels the divergent contributions and 
removes completely the contribution of the corresponding heat kernel 
coefficients. In the first two models we have:
\beq
p &\to &p \mp \frac{m^4} {64\pi^2}
\left( \frac 1 {s}
-\frac 1 2 +\ln \left[\frac{4\mu^2}{m^2}\right]\right),\quad
\sigma  \,\to  \, \sigma +\frac{m^3}{48\pi}, \nn\\
F  &\to &  F\pm \frac{m^2}{12\pi} 
\left( \frac 1 {s}
- 1  +\ln \left[\frac{4\mu^2}{m^2}\right]\right), \quad
k\, \to \, k-\frac m {96},
\label{n8}\\
h& \to & h\pm \frac 1 {630\pi}  
\left( \frac 1 {s}
- 2  +\ln \left[\frac{4\mu^2}{m^2}\right]\right),\nn
\eeq
where the upper sign corresponds to the first model and the lower sign to the
second.
In the third model there are only two  contributions,
 which are divergent in some regularizations, and the renormalization reads:
\beq
\sigma  \to  \sigma +\frac{m^3}{24\pi}, \quad 
k\to k-\frac{m}{48}. \label{n8a}
\eeq
Within the zeta functional regularization this is a finite renormalization. 

After the subtraction of these contributions from $E_0$ we
 denote it by $E_0^{(ren)}=E_0-E_0^{(div)}$ and the complete energy becomes
\beq
E=E_{(class)}+E_0^{(ren)}.
\eeq
Within our renormalization scheme, we have defined a unique renormalized 
groundstate energy $E_0^{(ren)}$. Often, the renormalization arbitrariness
is removed by imposing some normalization conditions. In this case a 
natural candidate would be the requirement, that $E_0^{(ren)}$ vanishes for 
$R\to\infty$. This is fulfilled in our case.
However, this requirement does not fix completely the renormalization in the first
two models because a finite renormalization of $h$ is still possible.

With respect to this renormalization there is a qualitative difference between
the first two and the third model. In the latter one some divergences have
been cancelled  when adding up the contributions from the interior and exterior
regions. The corresponding terms 
in the classical energy do not need to be 
renormalized. In general, as mentioned in \cite{andreas}
and much earlier in connection with the renormalization of QED, those 
contributions which need a renormalization are not of quantum nature but have 
to be present in the classical part of the system and are to be determined 
from outside (like the electron mass and charge in QED) or by the dynamics of the 
classical part. For example, in the bag model one has to look for a minimum of 
the complete energy while varying these parameters. The contributions which 
do not need renormalization, like the one resulting from $B_0$, $B_1$ or $B_2$ 
in the third model, may be absent in the classical part and can be considered 
as pure quantum contributions. In this sense the third model contains only two 
classical parameters ($\sigma$ and $k$). 

Having in mind that only the complete energy has 
a physical meaning, a change of
the normalization conditions, respectively a finite renormalization, would be 
equivalent
to a change of the classical parameters and should not influence issues like 
that of
finding a minimum of the complete energy by varying the parameters of the 
classical part. 

Special remarks are in order for the case of a massless quantum field. There
only contributions from $B_2$ remain (one performs the limit $R\to 0$ in the
regularized expressions, the $B_i$'s are proportional to $m^{4-2i}$). In the
third model these contributions are finite and can be considered as pure
quantum corrections. They yield, together with the finite contributions
(cf. the quantity $N$ in the next section), the known $1/R$ contributions to the
Casimir energy for a sphere and a massless field with various boundary
conditions. 
However, in the models (i) and (ii) the $B_2$ contribution is divergent and
the corresponding $1/R$ term in the energy must be considered as a classical
contribution. Thus, in these cases the ground state energy for a massless field
can be removed by a finite renormalization and the energy of the system 
remains formally the classical one. In that sense there is no Casimir effect.
The same is true in the presence of a thick spherical shell (interior and 
exterior region are separated by a finite distance with no quantum field),
because here no cancellation between interior and exterior modes occurs.

Similar remarks hold not only for a spherical bag but also for an arbitrarily 
shaped bag. For the infinitely thin bag a cancellation between interior
and exterior modes does occur, while if it has
 a finite thickness this is not true anymore.
 
\section{Calculation of the ground state energy}
First we consider the interior case. As it is easily  seen from equation 
(\ref{n3}), the task that remains for the
evaluation of the zero-point energy is to perform a convenient analytical
treatment of
the zeta function (\ref{intzeta}).
A precise way to obtain an analytic continuation of $\zeta (s)$
to $s=-1/2$ has been described in \cite{bk,bek} in  detail, what
allows us to be brief here.
We may write the zeta function for the interior space in the form
\beq
\zeta_{(int)} (s) =N_{(int)}(s) +\sum_{i=-1}^3 A_i (s),\label{4}
\eeq
where the $A_i$'s are the contributions of the first five terms of the uniform
asymptotic expansion of the modified Bessel functions as $\nu \to \infty$ and $k
\to \infty$, with $\nu /k$ fixed. It is sufficient to subtract these five
contributions in order to absorb all possible divergent contributions. A higher 
number of
subtractions is possible in order to speed up the convergence of the remaining
numerical expressions.  We have called $N$  
the zeta function where all these asymptotic terms
have been subtracted:
\beq
N_{(int)}(s)&=&\frac{\sin (\pi s)} {\pi} \sum_{l=0}^{\infty} (2l+1)
\int\limits_{mR/\nu}^{\infty}dx\,\, \left[\left(\frac{x\nu}R\right)^2-m^2
\right]^{-s }\nn\\
& &\times\frac{\partial}{\partial x}\left[\ln I_{\nu} (\nu x)-\ln \left(
\frac{e^{\nu\eta}}{\sqrt{2\pi\nu}(1+x^2)^{1/4}}\right)\right.
\label{5}\\
& &
\left. -\frac 1 {\nu} D_1 (t) -
\frac 1 {\nu^2} D_2 (t) -\frac 1 {\nu^3}
D_3 (t) \right],
\nn
\eeq
where we have set $t=1/\sqrt{1+x^2}$, and $\eta = \sqrt{1+x^2}
+\ln [x/(1+\sqrt{1+x^2})]$. 
In this formula the parameter $s$ can be put equal to $-1/2$  under
 the  integration and summation signs. The evaluation of 
$N(1/2)$ is the remaining numerical task.
The polynomials $D_i$ are
\beq
D_1 (t) &=& \sum_{a=0}^1 x_{1,a} t^{1+2a}\equiv
\frac 1 8 t -\frac 5 {24}t^3,\nn\\
D_2 (t) &=& \sum_{a=0}^2 x_{2,a} t^{2+2a}\equiv
\frac 1 {16} t^2 -\frac 3 8 t^4 +\frac 5 {16} t^6,\label{6}\\
D_3 (t) &=& \sum_{a=0}^3 x_{3,a} t^{3+2a}\equiv
\frac{25}{384} t^3 -\frac{531}{640} t^5
+\frac{221}{128} t^7 -\frac{1105}{1152}t^9, \nn
\eeq
and, in terms of their coefficients, $x_{i,a}$, the functions $A_i(s)$
are given by
\beq
A_{-1} (s) &=&\frac{\rzs}{2\sqrt{\pi}\Gamma (s)}\sujnu \lll
(mR)^{2j}\frac{\g j+s-\frac 1 2\right)}{s+j}\zeta_H (2j+2s
-2;1/2),\nn\\          
A_0 (s) &=&-\frac{\rzs}{2\Gamma (s)}\sujnu \lll (mR)^{2j}\Gamma (s+j)
\zeta_H (2j+2s-1;1/2),\label{7}
\\
A_i (s) &=& -\frac{2\rzs}{\Gamma (s)}\sujnu \lll (mR)^{2j} \zeta_H
(-1+i+2j+2s;1/2) \nn\\
& &\hspace{3cm} \times \suani x_{i,a} \frac{\g s+a+j+\frac i
2\right)}{\g a+\frac i 2\right)}.\nn
\eeq
It is easy to see that the above series are convergent for $|mR|
\leq 1$. Alternative representations valid for arbitrary values of
$mR$ are derived in App. A.
Using the above formulas, or alternatively, the formulas given in the
appendix, one can perform the renomalization and calculate the renormalized
ground state energy numerically. The result for the 
ground state energy is shown in Fig. 1 for $R=1$ as
a function of $m$.

\begin{figure}
\centerline{\psfig{figure=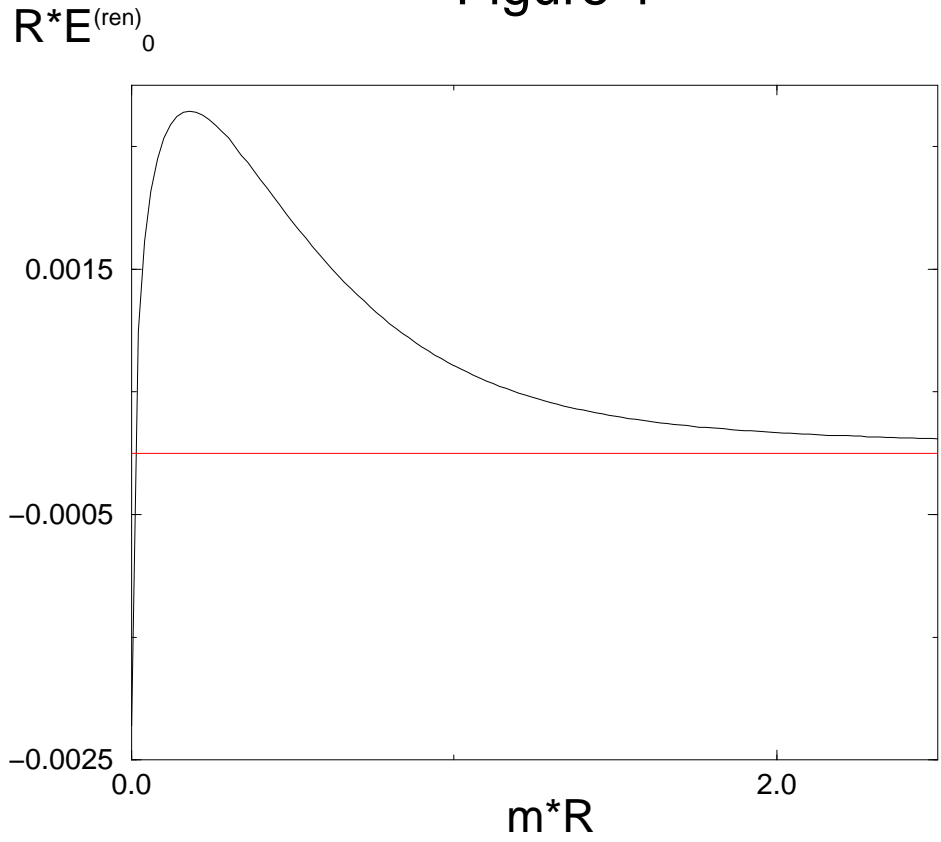,height=12cm}}
\caption{Plot of the renormalized vacuum energy
$E^{ren}_0$ measured in units of the inverse of the radius.}
\label{interior}
\end{figure}

The dependence of $E^{ren}_0$ on the radius for fixed
mass is depicted in Fig. 2.
It is seen that there is a maximimum for $m\,R \simeq 0.023$.
For very small values of the argument $m\,R$, the figure diverges to
$-\infty $, whereas for large values, the function goes to zero. In Fig. 2
we have restricted the domain around the aforementioned maximum.

\begin{figure}
\centerline{\psfig{figure=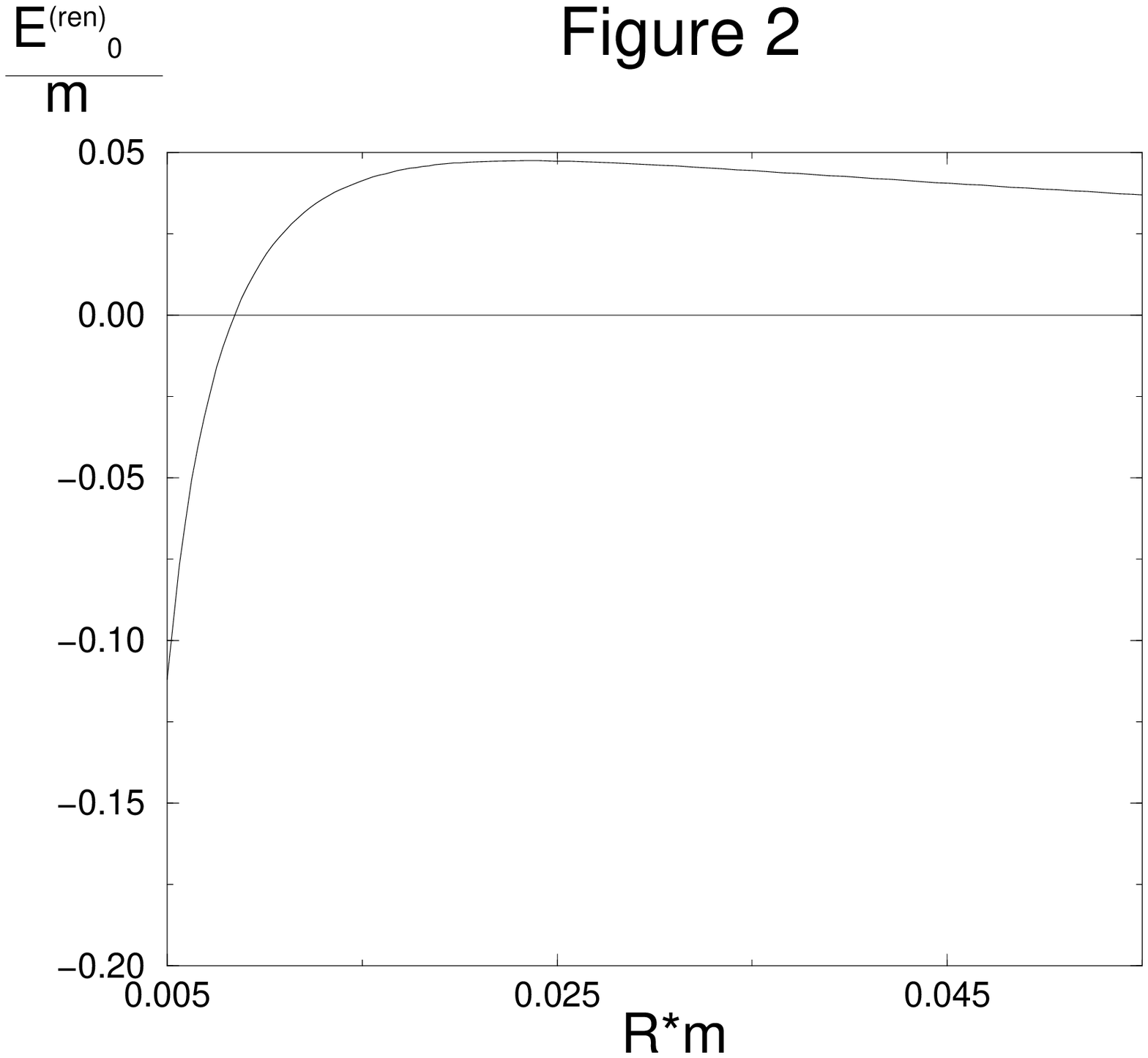,height=12cm}}
\caption{Plot of the renormalized vacuum energy
$E^{ren}_0$ measured in units of the mass. The plot has
been restricted to a domain around the maximum value.}
\label{interiortrn}
\end{figure}

The zero-point energy in the exterior of the spherical bag can be
calculated in a much similar manner.
Indeed, only a few changes are necessary.
Subtracting the Minkowski space zeta function from the zeta function 
associated with the field outside the bag, the starting point here reads
\beq
\zeta_{ext} (s) =N_{ext}(s) +\sum_{i=-1}^3 (-1)^i A_i (s),\label{8}
\eeq
with
\beq
N_{ext}(s)&=&\frac{\sin (\pi s)} {\pi} \sum_{l=0}^{\infty} (2l+1)
\int\limits_{mR/\nu}^{\infty}dx\,\, \left[\left(\frac{x\nu}R\right)^2-m^2
\right]^{-s }\nn\\
& &\times\frac{\partial}{\partial x}\left[\ln K_{\nu} (\nu x)-\ln \left(
\frac{\sqrt{\pi}e^{-\nu\eta}}{\sqrt{2\nu}(1+x^2)^{1/4}}\right)\right.
\label{9}\\
& &
\left. +\frac 1 {\nu} D_1 (t) -
\frac 1 {\nu^2} D_2 (t) +\frac 1 {\nu^3}
D_3 (t) \right].
\nn
\eeq
As is clear, one just needs to substitute the Bessel function
 $K_{\nu}$ for $I_{\nu}$.
The asymptotic contributions, as compared with those for the interior space,
 get an alternating
sign coming from the asymptotics of the Bessel function $K_
{\nu}$, which exhibit this sign, as
compared with those of the function $I_{\nu}$ (see  Ref. \cite{abramo}).
By construction, $N_{ext} (s)$ is finite at $s=-1/2$. The results for
$A_i (s=-1/2)$ are given in App. A. They are the same as in the previous
case. Again, the renormalized ground state energy can be calculated. The result
is shown in Fig. 3.
It is apparent that the slope is always negative
and that the plot always gives positive values. 
It is clear that, had we plotted the same quantity in units of the mass, a
curve with both these properties would have been obtained too. In particular,
it would not have a maximum as the one observed for the interior case.

\begin{figure}
\centerline{\psfig{figure=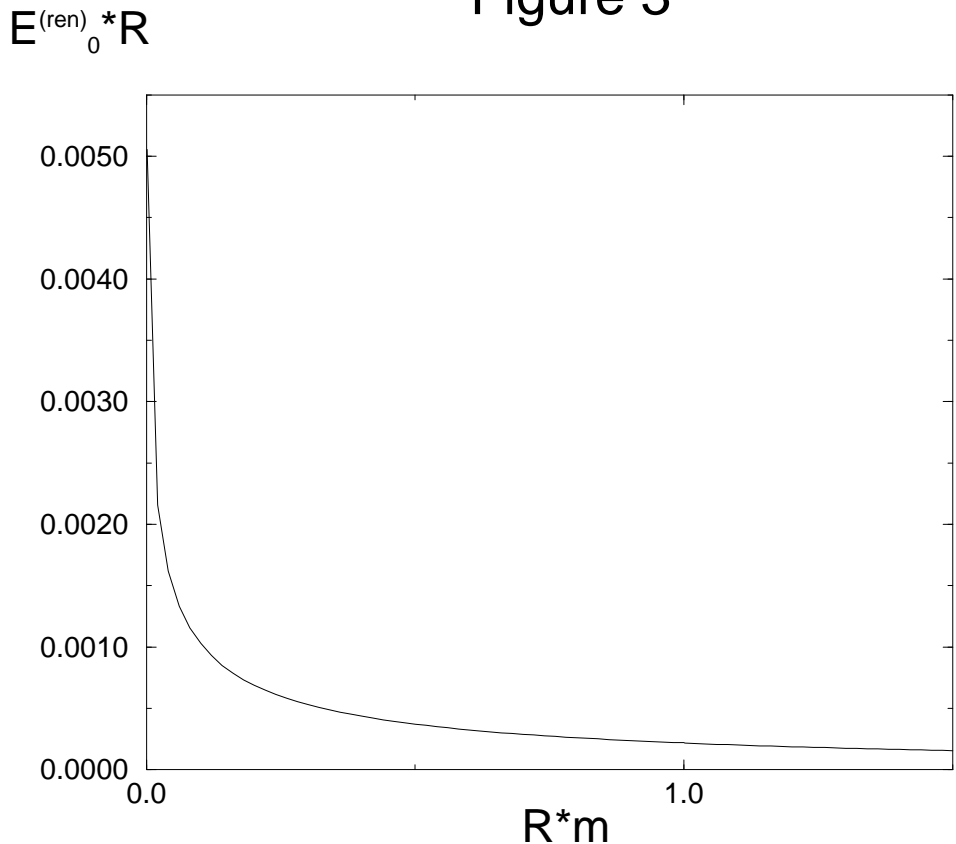,height=12cm}}
\caption{Plot of the renormalized vacuum energy in units of the
inverse of the radius. }
\label{exterior}
\end{figure}

In the case of the third model, i.e., for the quantum field extending to both regions 
altogether, we just have  to add the two results above. As is shown in Fig. 4,
there is an interval where the slope is positive. This would seem to leave open the
possibility that a plot in units of the mass could
 exhibit a maximum. We have carefully
investigated this possibility but the answer is negative. In other words,
such alternative plot is always monotonically decreasing. 

\section{Conclusions}
In this paper we have developed a systematic
 approach to the calculation of the
Casimir energy of a massive field in the presence of a spherical bag.
The models corresponding to the quantum field
 confined to the interior and to the exterior
space of the bag, respectively, have been discussed separately, 
and the differences in the renormalization of these models with respect 
to the case where the field is present in 
the whole space have been investigated in detail.

Figures 1-4 show the quantum contribution to the renormalized ground state energy. 
This quantum ground state energy of the 
interior region exhibits a maximum for 
variable radius and fixed mass, as is clear from Fig. 2.
Thus, we may say that if the bag is small enough,
the quantum part of the vacuum energy induces an
attractive force which is partly a result of our
renormalization prescription by which the energy is normalized to zero
in the infinite mass limit. This is the physically correct choice and, in order 
to deal with it, treatment of the massive case is necessary.  
The appearance of attractive forces in this kind of situation is not
new, however, and has been found also when dealing with spinors 
\cite{milton80} and with the electromagnetic field on a dielectric 
cylinder \cite{brevik}.

\begin{figure}
\centerline{\psfig{figure=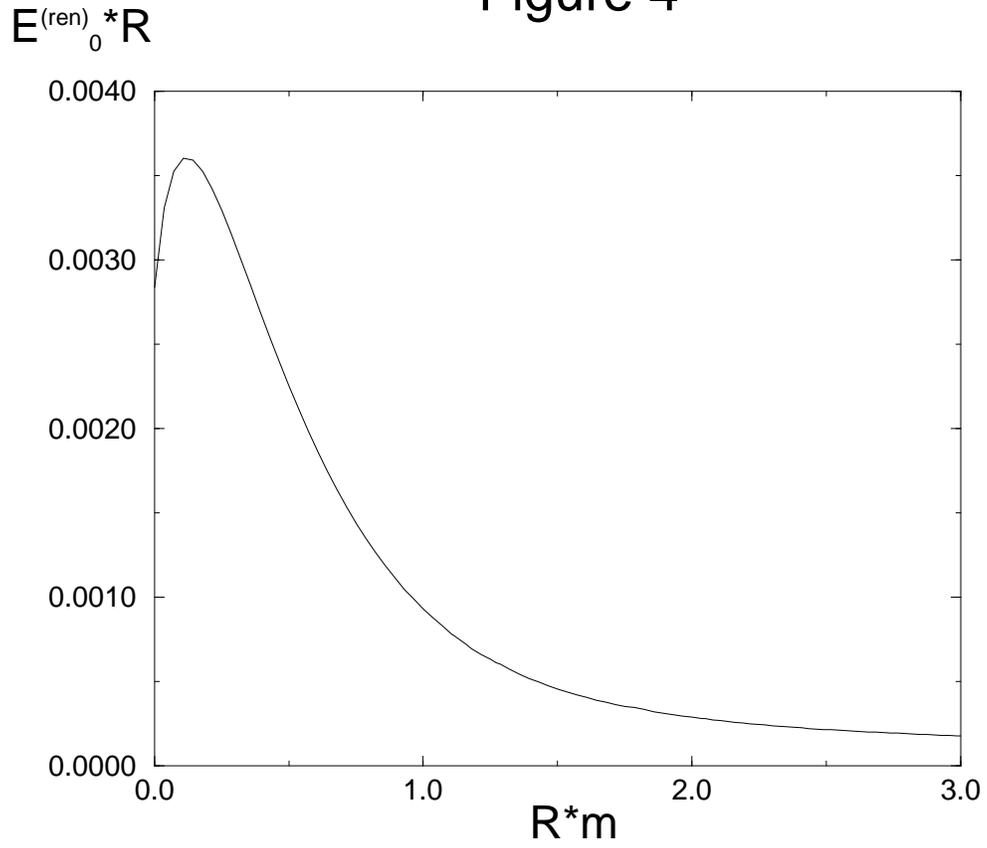,height=12cm}}
\caption{The renormalized vacuum energy  represented 
in units of the inverse of the radius.}
\label{tot}
\end{figure}

Robin boundary conditions can be treated in complete analogy, as 
has been described
in detail in Refs. \cite{bek}-\cite{bdk}. 
Also the interior and exterior regions can be considered separately, and
adding up the contributions coming from each region the same cancellation of
divergences appear. The ground state energy of the electromagnetic field 
is the sum of the ground state energy of two scalar fields satisfying,
respectively, 
Dirichlet boundary conditions and Robin boundary conditions (TE and TM modes).
The $B_{1/2}$ heat kernel coefficient has opposite sign for Dirichlet and
Robin boundary conditions, what leads to a partial cancellation of divergences
between  the TE and TM modes. Doing the same kind of calculation than the one
presented here and taking the massless limit, previous results are 
reobtained \cite{baldup78,mil78}, what provides a further check of the procedure.  
  
Along the same lines,
it would be
interesting to perform the calculations for 
higher spin fields and to apply the results
 to realistic physical models, as  the  MIT bag
model for instance. Furthermore, in complete analogy,  the case of 
two spherical shells can be treated with
 our method. Also,  the influence of dielectric media
can be considered for the case of an electromagnetic field.   
\section*{Acknowlegments}
This investigation has been supported by DGICYT (Spain), project PB93-0035,
by CIRIT (Generalitat de Catalunya), by the german-spanish program Acciones
Integradas, project HA1995-0171 and by DFG, contract Bo 1112/4-1. S.L. 
gratefully acknowledges an FI grant from Generalitat de Catalunya.
\appendix
\app{Appendix: Representations for the asymptotic contributions
inside the bag}
\setcounter{equation}{0}
In this appendix we derive explicit representations of the $A_i (s),
i=-1,...,3$
(see Eq. (\ref{7})), which are valid for arbitrary $mR$. Let us start with
$A_{-1} (s)$,  which is actually the most difficult piece to treat. Instead of
equation (\ref{7}), one may use the representation \cite{bk,bek}
\beq
A_{-1} (s) = 2\sip \sulnu \nu^2 
\int\limits_{mR/\nu}^{\infty} \left[\left(\frac{x\nu}R\right)^2-m^2
\right]^{-s} \frac{\sqrt{1+x^2}-1} x, \label{a1}
\eeq
of which we need the analytical continuation to $s=-1/2$.
With the substitution $t=(x\nu /R)^2 -m^2$, this expression results into
the following one
\beq
A_{-1} (s) 
&=& \sip \sulnu \nu \int\limits _0 ^{\infty}
dt\,\,\frac{t^{-s}}{t+m^2}\left\{
\sqrt{\nu^2 +R^2 (t+m^2)} -\nu\right\}\nn\\
&=&-\frac 1 {2\sqrt{\pi}}
 \sip \sulnu \nu \int\limits _0 ^{\infty}
dt\,\,t^{-s} \int\limits d\alpha \,\, e^{-\alpha (t+m^2)}\label{a2}\\
& &\times\int\limits_0^{\infty}d\beta \,\,\beta ^{-3/2} \left\{
e^{-\beta (\nu^2 +R^2 [t+m^2] ) } -e^{-\beta \nu^2}\right\},\nn
\eeq
where the Mellin integral representation for the single factors has been 
used. As we see,
the $\beta$-integral is well defined. Introducing a regularization parameter
$\delta$, $A_{-1}(s)$ can then be written as
\beq
A_{-1} (s) =\lim_{\delta \to 0} \left[ A_{-1}^1 (s,\delta )
+A_{-1} ^2 (s,\delta ) \right], \label{a3}
\eeq
with
\beq
A_{-1}^1 (s,\delta ) 
 &=&-\frac 1 {2\sqrt{\pi}}
 \sip \sulnu \nu \int\limits d\alpha \,\,
 e^{-\alpha m^2}
\int\limits_0^{\infty}d\beta \,\,\beta ^{-3/2+\delta}e^{-\beta (\nu^2 +R^2 
m^2 ) }
\int\limits _0 ^{\infty}
dt\,\,t^{-s}e^{-t(\alpha +\beta R^2)}\nn
\eeq
and
\beq
A_{-1} ^2 (s,\delta ) =    \frac 1 {2\sqrt{\pi}} \Gamma (1-s)
\sip    \sulnu \nu
\int\limits d\alpha \,\,
e^{-\alpha m^2 }   \alpha^{s-1}\int\limits_0^{\infty}d\beta \,\,
\beta ^{-3/2+\delta}e^{-\beta \nu^2}.
\nn
\eeq
Let us proceed with the remaining pieces. In $A_{-1}^1 (s,\delta )$
 one of the integrals can be done, yielding
\beq
A_{-1}^1 (s,\delta )
&=& -\frac{R^{1-2\delta}}{2\sqrt{\pi}\Gamma (s)} \Gamma (s+\delta -1/2)
\times\label{a4}\\
& &
\sulnu \nu \int\limits  _0 ^{\infty}
dy\,\, y^{\delta -3/2} \left[ m^2 +y \left(\frac{\nu} R\right)^2 
\right]^{1/2-s-\delta}. \nn
\eeq
For $A_{-1} ^2 (s,\delta )$, one gets
\beq
A_{-1} ^2 (s,\delta )&=& \frac{m^{-2s}}{2\sqrt{\pi}}\Gamma (\delta -1/2)
\sulnu \nu^{2-2\delta} \nn\\
&=& \frac{R^{1-2\delta}}{2\sqrt{\pi}\Gamma (s)}
\Gamma (s+\delta -1/2)
\sulnu \nu \int\limits  _0 ^{\infty}
dx\,\, x^{s-1        } 
\left[ m^2 x +
\left(\frac{\nu} R\right)^2 \right]^{1/2-s-\delta}.
\label{a5}
\eeq
And adding up (\ref{a4}) and (\ref{a5}) yields
\beq
A_{-1} (s) = \frac R {2\sqrt{\pi}\Gamma (s)} 
\Gamma (s -1/2) \sulnu \nu \int\limits  _0 ^{1}
dx\,\, x^{s-1        } 
\left[ m^2 x +
\left(\frac{\nu} R\right)^2 \right]^{1/2-s}, \label{a6}
\eeq
a form suited for the treatment of the angular momentum sum.

To perform the summation over $l$, we will use
\beq
\sum _{\nu=1/2} ^{\infty} f( \nu ) =\int_0^\infty d\nu \,\, f(\nu ) -i
\int_0^{\infty}d\nu \,\, \frac{f(i\nu +\epsilon )- f(-i\nu +\epsilon )}
{1+e^{2\pi \nu}},\label{a7}
\eeq
with $\epsilon\to 0$ understood. In order to get the Casimir energy we will 
need only the expansion of $A_{-1} (s)$ around $s=-1/2$. Using (\ref{a7})
 one  finds, after a lenghty
calculation,
\beq
A_{-1}(s) &=&\pol\left(\frac 7 {1920\pi R} +\frac{m^2R}{48\pi}
-\frac{m^4R^3}{24\pi}\right)\nn\\
& &+\ln4\left(\frac 7 {1920\pi R} +\frac{m^2R}{48\pi} 
-\frac{m^4R^3}{24\pi}\right)\label{a8}\\
& &+\frac 7 {1920\pi R} -\frac{m^2R}{48\pi}+\frac{m^4R^3}{48\pi}
\left(1+4\ln (mR)\right)\nn\\
& &-\frac 1 {\pi R} \inu \frac{\nu}{\nen} (\nr )\ln|\nr |\nn\\
& &-\frac{2m^2R}{\pi} \inu \frac{\nu}{\nen}\left(\ln |\nr |
+\frac{\nu}{mR}\ln\left|\frac{mR+\nu}{mR-\nu}\right|\right).\nn
\eeq
All other $A_i (s)$ can be treated in a much easier way. As a starting point
for $A_0 (s)$ we choose \cite{bk,bek}
\beq
A_0 (s) &=& -\frac{m^{-2s}} 2 \sum_{l=0}^{\infty} 
\nu \left[ 1+\left( \frac{\nu}{mR}\right)^2 \right]^{-2s}.
\label{a9}
\eeq
Using eq. (\ref{a7}), this yields immediately
\beq
A_0(s)&=&\frac 1 6 R^2m^3 -m\int\limits_0^{mR}d\nu\,\,
 \frac{\nu}{\nen}\sqrt{1-\numr}.\label{a10}
\eeq
For the remaining $A_i $'s we proceed in a different way. Let us explain
the method using one of the contributions of $A_1 (s)$,  say:
\beq
\sum_{l=0}^{\infty} 
\nu^{-2s}
\left[ 1+\left( \frac{\nu}{mR}\right)^2 \right]^{-s-1/2}
&=&\zeta_H (2s;1/2) -(s+1/2) (mR)^2 \zeta_H (2s+2;1/2) \nn\\
& & \hspace{-15mm} +\sum_{l=0}^{\infty}
\nu^{-2s}\left\{
\left[ 1+\left( \frac{\nu}{mR}\right)^2 \right]^{-s-1/2}
-1+(s+1/2) \left( \frac{mR}{\nu}\right)^2\right\}.
\nn
\eeq
This provides the immediate continuation of the sums to $s=-1/2$. In fact, the 
infinite sum is convergent and in the Hurwitz zeta function the 
analytical continuation to $s=-1/2$ is easily performed.
All pieces in $A_i$, $i=1,2,3$ have a similar aspect and may be treated
in the same way. Thus (we just write down the results)
\beq
A_1 (s) &=& \pol \left(\frac 7 {48 \pi }m^2 R +\frac 1 {192 \pi R}\right)\nn\\
& &-\frac 1 {72 \pi R} \left( 2+9\zeta_R' (-1)\right)
+\frac 1 {24 \pi} m^2 R (-2+7\gamma +21 \ln 2 )\nn\\
& &+\frac 1 {8\pi R}\sulnu \nu \left[\mrnu -\ln\left( 1+\mrnu \right) \right]
\nn\\
& &-\frac 5 {12\pi R} \sulnu \nu \left[\left( 1+\mrnu \right) ^{-1}
-1+\mrnu \right],\nn\\
A_2 (s) &=& \frac 1 {16 R} \sulnu \left[\left(
 1+\mrnu \right) ^ {-1/2} -1\right]
-\frac 3 {16R} \sulnu \left[ \left( 1+\mrnu \right)^{-3/2}-1\right]
\nn\\
& &+\frac{15}{128R} \sulnu \left[ \left(1+\mrnu \right) ^{-5/2} -1\right],
\label{a11}\\
A_3 (s) &=& -\frac{229}{40320\pi R} \pol
+\frac{2152 -687\gamma -2061 \ln 2}{60480 \pi R}\nn\\
& &+\frac{25}{192\pi R}\sulnu \frac 1 {\nu}\left[\left(
1+\mrnu \right)^{-1}-1\right]
-\frac{177}{160\pi R}\sulnu \frac 1 {\nu}\left[\left(
 1+\mrnu \right)^{-2}-1\right]\nn\\
& &+\frac{221}{120\pi R}\sulnu \frac 1 {\nu}\left[\left(
 1+\mrnu \right)^{-3}-1\right]
-\frac{221}{252\pi R}\sulnu \frac 1 {\nu}\left[\left(
 1+\mrnu \right)^{-4}-1\right].\nn
\eeq
This completes the list of expressions necessary for the analysis of the 
massive scalar field inside the bag with Dirichlet boundary conditions.


\newpage

\begin{thebibliography}{10}

\bibitem{casimir56}
H.B.G. Casimir, 
Physica {\bf 19} (1953) 846.

\bibitem{casimir48}
H.B.G. Casimir, Proc. Koninkl. Ned. Akad. Wetenshap {\bf 51} (1948) 793.

\bibitem{boyer68}
T.H. Boyer, Phys. Rev. {\bf 174} (1968) 1764.

\bibitem{baldup78}
R. Balian and R. Duplantier, Ann. Phys. {\bf 112} (1978) 165.

\bibitem{mil78}
K.A. Milton, L.L. De Raad Jr. and J. Schwinger, Ann. Phys. {\bf 115} (1978) 388.

\bibitem{eorbz}
E. Elizalde, S.D. Odintsov, A. Romeo, A.A. Bytsenko and S. Zerbini.
{\it Zeta regularization techniques with applications},
World Sci., Singapore (1994).

\bibitem{ee}
E. Elizalde, {\it Ten physical applications of spectral zeta functions},
Springer, Berlin (1995).

\bibitem{greiner}
G. Plunien, B. M\"uller and W. Greiner, Phys. Rep. {\bf 134} (1986) 87.

\bibitem{cho74}
A. Chodos, R.L. Jaffe. K. Johnson, C.B. Thorn and V. Weisskopf,
Phys. Rev. D {\bf 9} (1974) 3471.

\bibitem{cho74a}
A. Chodos, R.L. Jaffe. K. Johnson and C.B. Thorn,
Phys. Rev. D {\bf 10} (1974) 2599.

\bibitem{bender74}
C.M. Bender and P. Hays, Phys. Rev. D {\bf 14} (1976) 2622.

\bibitem{has78}
P. Hasenfratz and J. Kuti, Phys. Rep. {\bf 40C} (1978) 75.

\bibitem{vep90}
L. Vepstas and A.D. Jackson, Phys. Rep. {\bf 187} (1990) 109;
Nucl. Phys. {\bf A481} (1988) 668.

\bibitem{rho83}
M. Rho, A.S. Goldhaber and G.E. Brown, Phys. Rev. Lett. {\bf 51} (1983)
747.

\bibitem{brow79}
G.E. Brown and M. Rho, Phys. Lett. {\bf B82} (1979) 177.

\bibitem{brow84}
G.E. Brown, A.D. Jackson, M. Rho and V. Vento, Phys. Lett. {\bf B140} 
(1984) 285.

\bibitem{milton80}
K.A. Milton, Phys. Rev. D {\bf 22} (1980) 1441; {\it ibid} 1444;
Ann. Phys. {\bf 127} (1980) 49.

\bibitem{milton83}
K.A. Milton, Ann. Phys. {\bf 150} (1983) 432.

\bibitem{baacke83}
J. Baacke and Y. Igarashi, Phys. Rev. D {\bf 27} (1983) 460.

\bibitem{andreas}
S.K. Blau, M. Visser and A. Wipf, Nucl. Phys. B {\bf 310} (1988) 163.

\bibitem{bk}
M. Bordag and K. Kirsten, {\it Heat-kernel coefficients of the Laplace operator
in the 3-dimensional ball}, hep-th/9501064.

\bibitem{bek} 
M. Bordag, E. Elizalde and K. Kirsten, J. Math. Phys. {\bf 37} (1996) 895. 

\bibitem{begk}
M. Bordag, E. Elizalde, B. Geyer and K. Kirsten, Commun. Math. Phys. {\bf 179}
(1996) 215.

\bibitem{bdk}
M. Bordag, S. Dowker and K. Kirsten, {\it Heat-kernels and functional
determinants on the generalized cone}, to appear in Commun. Math. Phys.

\bibitem{abdk}
J. Apps, M. Bordag, S. Dowker and K. Kirsten, {\it Spectral invariants for the
Dirac equations on the d-ball with various boundary conditions}, hep-th/9511060.

\bibitem{cg}
G. Cognola and K. Kirsten, Class. Quantum Grav. {\bf 13} (1996) 633.

\bibitem{eli1}  
E. Elizalde, M. Lygren and D.V. Vassilevich,
{\it Antisymmetric tensor fields on spheres: functional determinants
and non-local counterterms},
to appear in J. Math. Phys.

\bibitem{eli2}
E. Elizalde, M. Lygren and D.V. Vassilevich,
{\it Zeta function for the Laplace operator acting on forms in a ball
with gauge boundary conditions},
UB-ECM-PF 96/7, hep-th/9605026.

\bibitem{dowker}
J.S. Dowker and J.S. Apps, Class. Quantum Grav. {\bf 12} (1995) 1363; 
J.S. Dowker, Class. Quantum Grav. {\bf 13} (1996) 1; Phys. Lett. B 
{\bf 366} (1996) 89.

\bibitem{rom}
A. Romeo, Phys. Rev. D {\bf 52} (1995) 7308, {\bf 53} (1996) 3392;
S. Leseduarte and A. Romeo, Europhys. Lett. {\bf 34} (1996) 79-83;
hep-th/9605022,  to appear in Ann. Phys.

\bibitem{borkir}
M. Bordag and K. Kirsten, Phys. Rev. D {\bf 53} (1996) 5753.

\bibitem{kennedy78}
G. Kennedy, J. Phys. A: Math. Gen. {\bf 11} (1978) L173.

\bibitem{abramo}
M. Abramowitz and I.A. Stegun,
{\it Handbook of Mathematical Functions
  (Natl.~Bur.~Stand.~Appl.~Math.~Ser.55)},
(Washington, D.C.: US GPO),  Dover, New York, reprinted 1972.

\bibitem{brevik}
I. Brevik and G.H. Nyland, Ann. Phys. {\bf 230} (1994) 321.

\end{thebibliography}

\end{document}